\documentclass[twocolumn,preprintnumbers,showpacs,nofootinbib,floatfix]{revtex4}
\usepackage{graphicx,latexsym,amsmath,amssymb}
\usepackage{dcolumn}
\usepackage{bm}
\usepackage{longtable}%
\begin{document}
%
\newcommand{\Expt}  {Ex\-peri\-ment}
\newcommand{\Expts} {Ex\-peri\-ments}
\newcommand{\Expl}  {Ex\-peri\-men\-tal}
\newcommand{\expt}  {ex\-peri\-ment}
\newcommand{\expts} {ex\-peri\-ments}
\newcommand{\expl}  {ex\-peri\-men\-tal}
\newcommand{\spcF}  {spectroscopic factor}
\newcommand{\spcFs} {spectroscopic factors}
\newcommand{\sglp}  {single-par\-ticle}
\newcommand{\sglh}  {single-hole}
\newcommand{\ph}    {par\-ticle-hole}
\newcommand{\Ph}    {Par\-ticle-hole}
\newcommand{\coef}  {coefficient}
\newcommand{\coefs} {coefficients}
\newcommand{\Cfg}   {Con\-figu\-ration}
\newcommand{\cfg}   {con\-figu\-ration}
\newcommand{\cfgs}  {con\-figu\-rations}
\newcommand{\ONrule}{ortho-normality rule}
\newcommand{\angD}  {angular dis\-tri\-bution}
\newcommand{\angDs} {angular dis\-tri\-butions}
\newcommand{\AngDs} {Angular dis\-tri\-butions}
\newcommand{\excF}  {excitation function}
\newcommand{\excFs}  {excitation functions}
\newcommand{\ExcFs}  {Excitation functions}
\newcommand{\peneTra} {pene\-tra\-bi\-li\-ty}
\newcommand{\peneTras} {pene\-tra\-bi\-li\-ties}
\newcommand{\Cx} {\mbox{$^{12}$C}}
\newcommand{\Cy} {\mbox{$^{13}$C}}
\newcommand{\Cz} {\mbox{$^{14}$C}}
\newcommand{\Nx} {\mbox{$^{14}$N}}
\newcommand{\Ny} {\mbox{$^{15}$N}}
\newcommand{\Ox} {\mbox{$^{16}$O}}
\newcommand{\Oy} {\mbox{$^{17}$O}}
\newcommand{\Nax} {\mbox{$^{23}$Na}}
\newcommand{\ClF} {\mbox{$^{35}$Cl}}
\newcommand{\ClS} {\mbox{$^{37}$Cl}}
\newcommand{\Ce} {\mbox{$^{140}$Ce}}
\newcommand{\Sm} {\mbox{$^{147}$Sm}}
\newcommand{\Bi} {\mbox{$^{209}$Bi}}
\newcommand{\BiE} {\mbox{$^{208}$Bi}}   
\newcommand{\BiS} {\mbox{$^{207}$Bi}}   
\newcommand{\Pb} {\mbox{$^{208}$Pb}}
\newcommand{\PbF} {\mbox{$^{204}$Pb}}   
\newcommand{\PbX} {\mbox{$^{206}$Pb}}   
\newcommand{\PbS} {\mbox{$^{207}$Pb}}   
\newcommand{\PbN} {\mbox{$^{209}$Pb}}   
\newcommand{\PbT} {\mbox{$^{210}$Pb}}   
\newcommand{\Tl} {\mbox{$^{207}$Tl}}
\newcommand{\TlE} {\mbox{$^{208}$Tl}}   
\newcommand{\Ex}  {E\mbox{${_x}$}}	
\newcommand{\Ep}  {E\mbox{${_p}$}}	
\newcommand{\gam}  {\mbox{$\gamma$}}
\newcommand{\bet}  {\mbox{$\beta$}}
\newcommand{\bgam}  {\mbox{$\beta$--$\gamma$}}
\newcommand{\dpgam}  {\mbox{(d, p$\gamma$)}}
\newcommand{\ppgam}  {\mbox{(p, p'$\gamma$)}}
\newcommand{\ggam}  {\mbox{$\gamma$--$\gamma$}}
\newcommand{\pps}     {(p,p{\mbox{$'$)}}}
\newcommand{\talpha}  {(t,{\mbox{$\alpha$)}}}
\newcommand{\dHe}  {(d, {\mbox{$^3$He)}}}
\newcommand{\sOhlb} {s{\mbox{$_{{1}/{2}}$}}}   
\newcommand{\pOhlb} {p{\mbox{$_{{1}/{2}}$}}}   
\newcommand{\pThlb} {p{\mbox{$_{{3}/{2}}$}}}   
\newcommand{\dThlb} {d{\mbox{$_{{3}/{2}}$}}}   
\newcommand{\dFhlb} {d{\mbox{$_{{5}/{2}}$}}}   
\newcommand{\fFhlb} {f{\mbox{$_{{5}/{2}}$}}}   
\newcommand{\fShlb} {f{\mbox{$_{{7}/{2}}$}}}   
\newcommand{\gShlb} {g{\mbox{$_{{7}/{2}}$}}}   
\newcommand{\gNhlb} {g{\mbox{$_{{9}/{2}}$}}}   
\newcommand{\hNhlb} {h{\mbox{$_{{9}/{2}}$}}}   
\newcommand{\hEhlb} {h{\mbox{$_{{11}/{2}}$}}}  
\newcommand{\iEhlb} {i{\mbox{$_{{11}/{2}}$}}}  
\newcommand{\iThlb} {i{\mbox{$_{{13}/{2}}$}}}  
\newcommand{\jThlb} {j{\mbox{$_{{13}/{2}}$}}}  
\newcommand{\jFhlb} {j{\mbox{$_{{15}/{2}}$}}}  
\newcommand{\kShlb} {k{\mbox{$_{{17}/{2}}$}}}  
\newcommand{\Ohlb}  {\mbox{${\frac{1}{2}}$}}       
\newcommand{\Thlb}  {\mbox{${\frac{10}{2}}$}}      
%
%
\newcommand{\apprsim} {\mbox{${< \atop \sim}$}}	
%
%
\title
{
On the mixing strength in the two lowest
 $0^{-}$ states in \Pb 
}
\author{A. Heusler
}
 \affiliation{%
 Max-Planck-Institut f\"ur Kernphysik, 
 D-69029 Heidelberg, Germany
}%
\email[correspondance to: ]{A.Heusler@mpi-hd.mpg.de}
 \author{
 G. Graw
 }
 \affiliation{%
 Department f\"ur Physik, Ludwig-Maximilian-Universit\"at M\"unchen, 
 D-85748 Garching, Germany
 }
 \author{
 R. Hertenberger
 }
 \affiliation{%
 Department f\"ur Physik, Ludwig-Maximilian-Universit\"at M\"unchen, 
 D-85748 Garching, Germany
 }
 \author{
 R. Kr\"ucken
 }
 \affiliation{%
 Physik Department E12, Technische Universit\"at M\"unchen, 
 D-85748 Garching, Germany
 }
 \author{
 F. Riess
 }
 \affiliation{%
 Department f\"ur Physik, Ludwig-Maximilian-Universit\"at M\"unchen, 
 D-85748 Garching, Germany
 }
 \author{
 H.-F. Wirth
 }
 \altaffiliation[now at ]{Technische Universit\"at M\"unchen, 
 D-85748 Garching, Germany}
 \affiliation{%
 Department f\"ur Physik, Ludwig-Maximilian-Universit\"at M\"unchen, 
 D-85748 Garching, Germany
 }
\author{
P. von Brentano
}
 \affiliation{%
 Institut f\"ur Kernphysik,
 Universit\"at zu K\"oln, 
 D-50937 K\"oln, Germany
 }
%
\begin{abstract}
With a resolution of 3\,keV, the two lowest $0^{-}$ states in \Pb\
are identified by measurements of the reaction \PbS{\it(d,~p)} with
the M\"unchen Q3D magnetic spectrograph in a region where the average
level spacing is 6\,keV.
Precise relative spectroscopic factors are determined.  Matrix
elements of the residual interaction among one-particle one-hole
\cfgs\ in a two-level scheme are derived for the two lowest $0^{-}$
states in \Pb.  The off-diagonal mixing strength is determined
as $105\pm10\,(experimental)\pm40\,(systematic)$\,keV.
Measurements of the reaction \Pb{\it(p,~p'\,}) via isobaric analog
resonances in \Bi\ support the structure information obtained.
%
\end{abstract}
\maketitle
     
  \section{Introduction
}

The nucleus \Pb\ offers the singular chance to study a two-level
scheme in the space of shell model \cfgs.
Below $E_x=6.1$\,MeV, only two $0^{-}$ states among about 120
one-particle one-hole \cfgs\ are expected from shell model
calculations \cite{2000Brwn,Rej1999}.  They are identified
\cite{Schr1997} but their structure is not known in detail.
With the average residual interaction known from \expt\
\cite{AB1973,Heu2006} they are predicted to consist essentially of the
two lowest \cfgs\ \sOhlb\pOhlb\ and \dFhlb\fFhlb, since the next \ph\
\cfg\ is ten times more distant than an average matrix element of the
residual interaction among one-particle one-hole \cfgs\ (m.e.).

We took spectra of the reaction \PbS{\it(d,~p)} at a resolution of
3\,keV \cite{P.i11} up to $E_x=8$\,MeV and identified the two
$0^{-}$ states in the region $E_x=5.2-5.7$\,MeV where the mean
level distance is 6\,keV.

Most of the low-lying states in \Pb\ are considered as excited states
created by the coupling of exactly one particle and one hole to the
ground state.  We postulate that each \ph\ state is completely
described as a mixture of a few \ph\ \cfgs.  The ground state of \PbS\
is assumed to be a pure \pOhlb\ neutron hole state in relation to the
ground state of \Pb.  In the \PbS{\it(d,~p)} reaction, the \ph\ states
in \Pb\ with spin $0^{-}$ are populated by $L=0$ transfer only,
whereas the $1^{-}$ states are populated by both $L=0$ and $L=2$
transfer.

For two spin $0^{-}$ and nine $1^{-}$ states below $E_x=6.5$\,MeV,
relative spectroscopic factors are measured.  Using the method of
Ref.~\cite{AB1973} and assuming the two lowest \cfgs\ to be almost
completely contained in the two lowest $0^{-}$ states, matrix elements
of the residual interaction between the $0^{-}$ \cfgs\ \sOhlb\pOhlb\
and \dFhlb\fFhlb\ are deduced.

Results of the inelastic proton scattering on \Pb\ via isobaric analog
resonances (IAR) in \Bi\ populating the two $0^{-}$ states and some
$1^{-}$ states \cite{1968WH02,P.i11} are discussed.

\begin{figure}[htb]
\caption[
\PbS{\it(d,~p)} spectra.139
]
{\label{dp_spectra.139}
(online: color)
\PbS{\it(d,~p)} spectrum taken at $\Theta=30^\circ$ for
$E_x=5.23 - 5.36$\,MeV.  The 5280 $0^{-}$ state (marked~$\bullet$) is
resolved from the two neighbors in 4-7\,keV distance.  It is displayed
on a logarithmic scale since the background is 1/2000 of the maximum
peak, but many levels with 1\% of the maximum are clearly resolved.
The drawn curves show the fit by the computer code GASPAN
\cite{Rie2005}, where the energies are taken from Table~\ref{tab.0m}
and only the centroid of all energies together and the peak heights
are varied. The widths and tails are interpolated from a table
generated by inspection of several strong, rather isolated peaks in
the whole spectrum covering about 1.2\,MeV.
A weak contamination line from \Nax\ is identified near
$E_x=5.31$\,MeV.
}
\resizebox{\hsize}{07.51cm}{
{\includegraphics[angle=00]{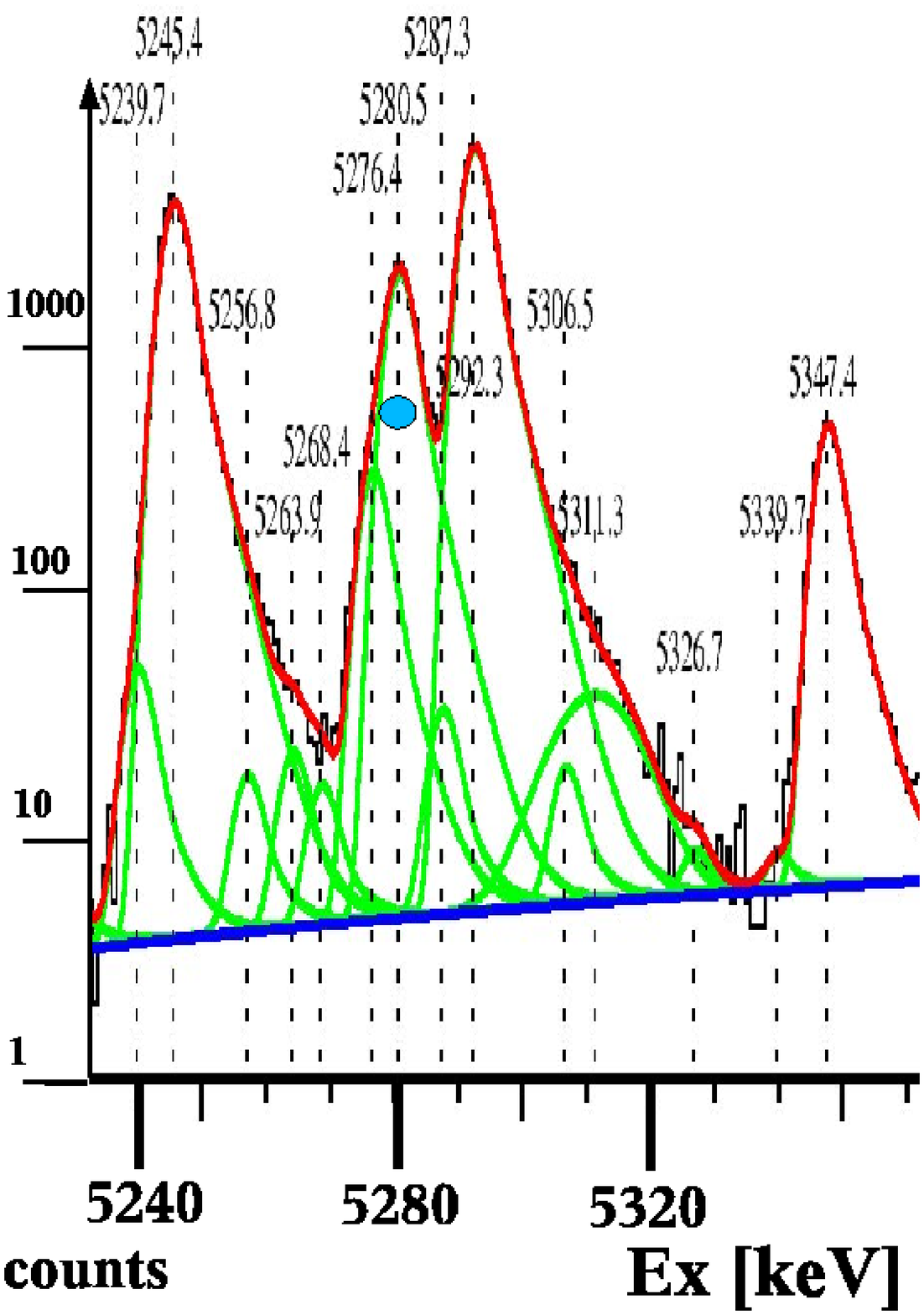}}
}
\end{figure}

\begin{figure}[htb]
\caption[
\PbS{\it(d,~p)} spectra.141
]
{\label{dp_spectra.141}
(online: color)
\PbS{\it(d,~p)} spectrum taken at $\Theta=25^\circ$ for $E_x=5.54-5.65$\,MeV.
The 5599 $0^{-}$ state (marked~$\bullet$) is well isolated.  For other
details see Fig.~\ref{dp_spectra.139}.
}
\resizebox{\hsize}{07.51cm}{
{\includegraphics[angle=00]{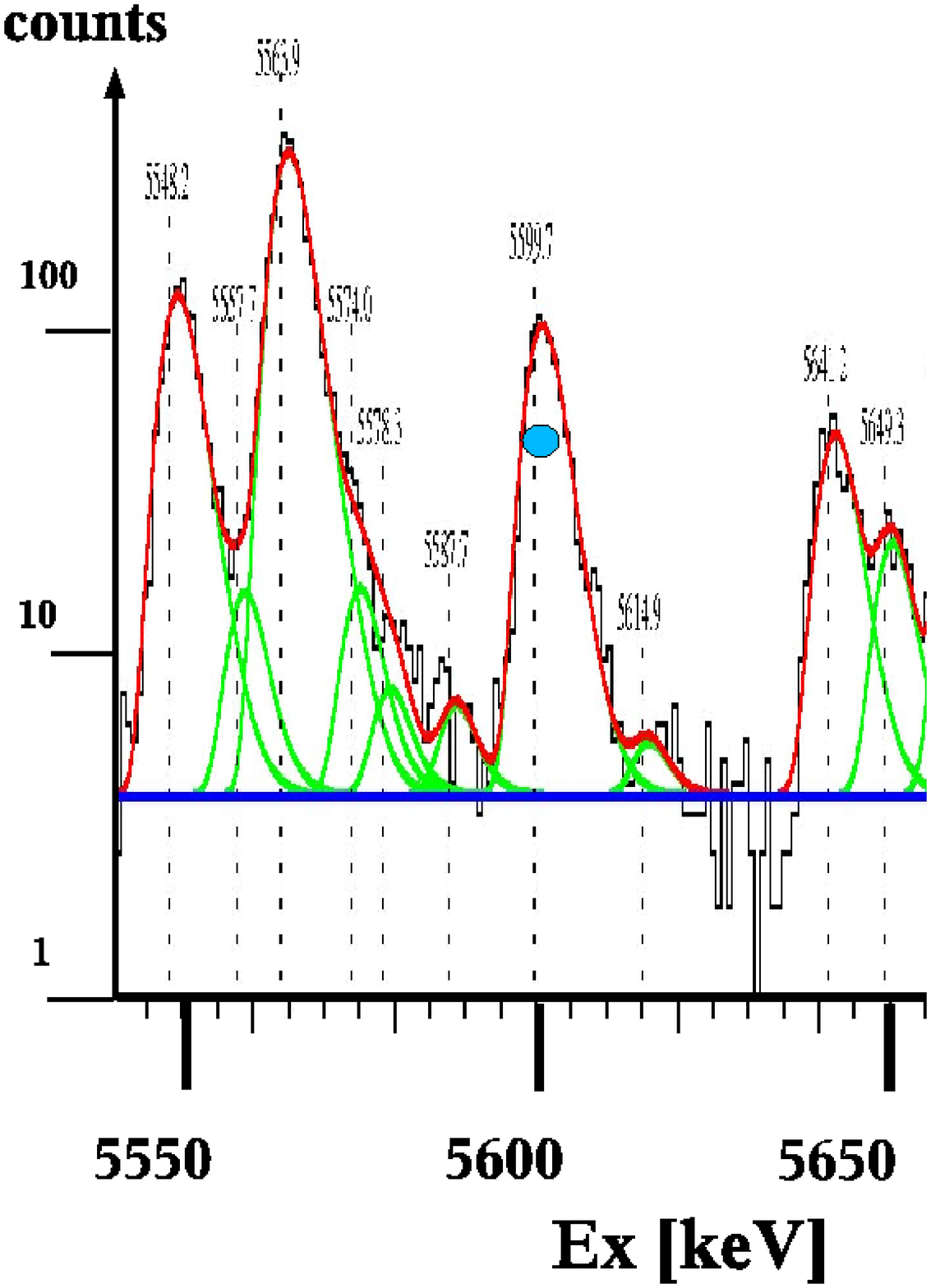}}
}
\end{figure}

\section{\Expt al data
}

\subsection{\Expt s with the Q3D magnetic spectrograph
}

Using the Q3D magnetic spectrograph of the tandem accelerator of the
Maier-Leibnitz laboratory at M\"unchen, \expt s of the reactions
\PbS{\it(d,~p)} and \Pb{\it(p,~p'\,}) via isobaric analog resonances
in \Bi\ (IAR-pp') are performed.  They are described in detail in
Ref.~\cite{P.i11}.  The resolution of about 3\,keV, the low background
(up to 1:5000) and reliable identification of contamination lines from
light nuclei (by the kinematic broadening proportional to different
slit openings), and a sophisticated fit of the spectra by the computer
code GASPAN \cite{Rie2005}, allow to resolve nearby levels and to
detect weakly excited states.  Here we refer to data obtained from the
\PbS{\it(d,~p)} \expt\ in the region $E_x=5.2-5.7$\,MeV.  
Compared to earlier work with a resolution of 18\,keV from the
Heidelberg multi-gap magnetic spectrograph \cite{Vol1973} and
following work \cite{NDS1986,Schr1997,Valn2001,ValnThesis}, the
resolution has been improved and the background lowered.

The mean level spacing is about 6\,keV in the regions near the two
$0^{-}$ states.  Peaks are identified by comparison to the known data
\cite{Rad1996,1997RA207,Yates96c,Schr1997,ValnThesis,Valn2001}, see
Table~\ref{tab.0m}.
A comparison to the preliminary analysis of the \Pb{\it(p,~p'\,}) data
on seven IAR in \Bi\ with similar resolution \cite{P.i11} allows to
verify the identifications.

Figs. \ref{dp_spectra.139} and \ref{dp_spectra.141} show two extracts
of \PbS{\it(d,~p)} spectra, each covering 1.2\,MeV totally.  Whereas
the neighbors of the 5599 $0^{-}$ state are 12-15\,keV away, the 5280
$0^{-}$ state is surrounded by two levels in 4-7\,keV distance.  At
scattering angles of $\Theta=20^\circ-30^\circ$, the 5276 and the 5287
state are excited with cross sections of 1-20\% of the 5280 state.

Peaks from light contaminations (\Cx, \Nx, \Ox, \Nax\ and more) are
identified in the whole spectra by the kinematic shift in a series of
spectra taken at scattering angles $\Theta=20^\circ- 30^\circ$ and the
kinematic broadening for different openings of the entrance slit to
the Q3D magnetic spectrograph, see Ref.~\cite{P.i11}.
In the region of $E_x = 5.5 - 5.7$\,MeV, contamination lines from \Nx\
with cross sections of a few $\mu$b/sr are detected at scattering
angles $\Theta=20^\circ$ and $30^\circ$.

\subsection{Extraction of relative spectroscopic factors
}

\begin{figure}[htb]
\caption[
S.F.
]
{\label{S.F.}
\AngDs\ of \PbS{\it(d,~p)}  for the 5280 $0^{-}$, 5292 $1^{-}$, 5599
$0^{-}$ states.  At $\Theta=20^\circ, 25^\circ, 30^\circ$ and for each
state, the mean cross section from six runs evaluated with
different methods of the background subtraction \cite{P.i11} is shown.
The cross section for the two $0^{-}$ states are reduced by the given
factors.
The dashed curve shows the DWBA calculation fitting the data of
Refs.~\cite{Valn2001,ValnThesis} for the 5292 $1^{-}$ state with
$L=0$.  For $L=2$ the DWBA curve and the data for the two levels (5924
$2^{-}$, 5947 $1^{-}$ \cite{Schr1997}) bearing the main strength of
the \dThlb\pOhlb\ \cfg\ vary by less than~10\% in between $\Theta=
20^\circ-30^\circ$.
}
\resizebox{\hsize}{11.81cm}{
{\includegraphics[angle=00]{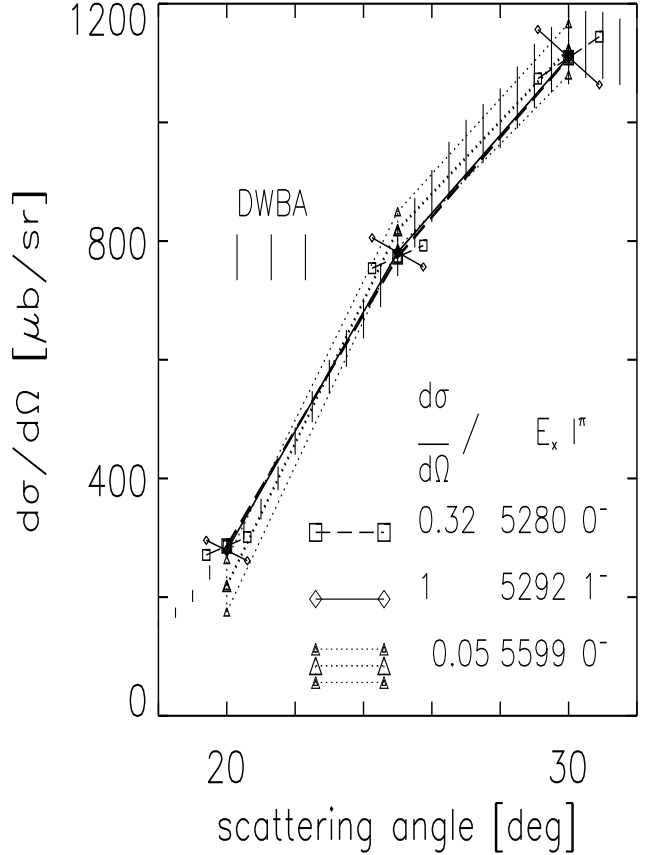}}
}
\end{figure}

By use of the GASPAN code \cite{Rie2005} with the option of fixed
energy distances, and the excitation energies from Table~\ref{tab.0m},
the cross sections are precisely determined.
Figs. \ref{dp_spectra.139} and \ref{dp_spectra.141} shows spectra for
the regions around the 5280 $0^{-}$ and the 5599 $0^{-}$
levels. Fig.~\ref{S.F.} shows the \angDs\ for the 5280 $0^{-}$, 5292
$1^{-}$ and 5599 $0^{-}$ levels.  For scattering angles $\Theta=
20^\circ- 30^\circ$, the cross sections differ by a constant factor
(0.32 and 0.05 for the two $0^{-}$ states in relation to 5292 $1^{-}$
state) within the errors.  For $\Theta=20^\circ-30^\circ$, DWBA
calculations yield the steep slope observed for $L=0$ in contrast to a
rather flat \angD\ for $L=2$ \cite{Valn2001,ValnThesis}.

In view of the weak cross sections at $\Theta=20^\circ$, especially
for the 5599 $0^{-}$ state, we determine relative spectroscopic
factors by first calculating a mean \angD\ of the three states,
\begin{eqnarray}
\label{eq.slope}
\widetilde{\frac{d\sigma}{d\Omega}}(\Theta)=
\sum_{E_x}{
\left\{
{\frac{d\sigma}{d\Omega}}(E_x,\Theta) 
/
\sum_\theta{\frac{d\sigma}{d\Omega}}(E_x,\Theta)
\right\}.
}
\end{eqnarray}
The energy dependence of the cross section is neglected because of the
small energy range.  In a least squares fit we then obtain the mean
cross section
\begin{eqnarray}
\label{eq.mean}
\left<{\frac{d\sigma}{d\Omega}}(E_x)
\right>
=
\sum_\theta{
\left\{
{\frac{d\sigma}{d\Omega}}(E_x,\Theta) /
\widetilde{\frac{d\sigma}{d\Omega}}(\Theta)
\right\}
}
\end{eqnarray}
as a measure of the relative spectroscopic factors. In
Table~\ref{tab.compare} we adjust the mean values to the cross section
of the 5292 state at the scattering angle $\Theta=25^\circ$.

\subsection{Determination of mixing amplitudes 
}

The lowest negative parity states in \Pb\ are assumed to be well
described by the shell model as \ph\ states in relation to the ground
state of \Pb.  Especially, the two lowest $0^{-}$ states $|E_x,I^\pi>$
are assumed to consist of the \cfgs\ $|\sOhlb\pOhlb> {\rm\ and\ }
|\dFhlb\fFhlb>$ with admixtures of higher \cfgs\ $|C_q>$,
\begin{eqnarray}
\label{eq.ampl.0m}
|5280,0^{-}> = t_{11} \,|\sOhlb\pOhlb> + t_{12} \,|\dFhlb\fFhlb> +
\nonumber\\
 \sum_q t_{1q}\,|C_q>,
\nonumber\\
|5599,0^{-}> = t_{21} \,|\sOhlb\pOhlb> + t_{22} \,|\dFhlb\fFhlb> +
\nonumber\\
 \sum_q t_{2q}\,|C_q>.
\end{eqnarray}
The \PbS{\it(d,~p)} reaction populates the \sOhlb\pOhlb\ component
only.

In contrast to spin $0^{-}$, for spin $1^{-}$ the shell model predicts
eight states below $E_x=6.5$\,MeV.
Two \cfgs, \sOhlb\pOhlb\ and \dThlb\pOhlb, of the identified $1^{-}$
states (Table~\ref{tab.compare}) are excited by the \PbS{\it(d,~p)}
reaction.  Hence the $n$ $1^{-}$ states are described by
\begin{eqnarray}
\label{eq.ampl.1m}
|n,1^{-}> = t_{1\, n 1} \,|\sOhlb\pOhlb> + t_{1\, n 3}
\,|\dThlb\pOhlb> +
\nonumber\\
\sum_q t_{1\,n q}\,|C_q>.
\end{eqnarray}

We want to determine the matrix elements of the residual interaction
between the two lowest $0^{-}$ \cfgs\ in \Pb.  In the truncated
two-level \cfg\ space of one-particle one-hole \cfgs, the matrix $t$
is only approximately unitary,
\begin{eqnarray}
\label{eq.unitary}
t t^\dagger=
\left(
\begin{array}{rr}
 1- d_{11} & d_{12}  \\
 d_{21} & 1- d_{22} \\
\end{array}
\right)
 \approx
\left(
\begin{array}{rr}
 1 & 0  \\
 0 & 1 \\
\end{array}
\right).
\end{eqnarray}
We postulate the deviation from unitarity to be small,
\begin{eqnarray}
\label{eq.devia}
d=
\left(
\begin{array}{rr}
 d_{11} & d_{12} \\
 d_{21} & d_{22} \\
\end{array}
\right)
 \approx 0.
\end{eqnarray}
Each element of the deviation matrix contains only products of the
amplitudes $t_{1q},t_{2q}$ of higher \cfgs\ assumed to be weak
[Eq.~(\ref{eq.ampl.0m})] and the amplitudes $t_{q1},t_{q2}$ of the
\cfgs\ \sOhlb\pOhlb, \dFhlb\fFhlb\ in higher excited states assumed to
be weak, too.

According to the shell model without residual interaction, the two
\cfgs\ \sOhlb\pOhlb\ and \dFhlb\fFhlb\ have the lowest excitation
energies for the $1^{-}$ states, too.  For the $1^{-}$ states a
similar deviation matrix can be defined with elements $d_{1\,n1}$,
$d_{1\,n2},\ n=1,9$ referring to these two \cfgs.

An essential assumption is the proportionality of the sum of the
strengths of the \cfg\ \sOhlb\pOhlb\ in all states for the spins
$I^\pi= 0^{-}, 1^{-}$ to the spin factor $(2I+1)$,
\begin{eqnarray}
\label{eq.spinFactor}
\sum_{n} t^2_{1\,n 1} = 3
(t^2_{11}+t^2_{21} + d_{11}).
\end{eqnarray}
We then use the observation that the \cfgs\ \sOhlb\pOhlb\ and
\dThlb\pOhlb\ produce \angDs\ which are easily distinguished, to
derive upper and lower limits of the complete \sOhlb\pOhlb\ strength
$\sum_{n} t^2_{1\,n 1}$ in the $1^{-}$ states and thus derive an upper
limit for the deviation matrix $|d|$ by use of
Eq.~\ref{eq.spinFactor}.

Since the reaction \PbS{\it(d,~p)} excites only the \sOhlb\pOhlb\
component of the $0^{-}$ states [Eq.~(\ref{eq.ampl.0m})], the ratio of
the measured mean cross sections (Table~\ref{tab.compare})
\begin{eqnarray}
\label{eq.ratio}
t^2_{21}/t^2_{11}=
\left<{\frac{d\sigma}{d\Omega}}(5599)\right>
/
\left<{\frac{d\sigma}{d\Omega}}(5280)\right>
\end{eqnarray}
is used to derive the amplitudes $t_{11},t_{12},t_{21},t_{22}$ as
\begin{eqnarray}
\label{eq.ampl}
|t_{11}|= |t_{22}|=0.928\pm0.012,
\nonumber\\
|t_{12}|=|t_{21}|=0.37\pm0.04.
\end{eqnarray}
Here the deviation matrix $d$ [Eq.~(\ref{eq.devia})] is assumed to
vanish.

\begin{figure}[htb]
\caption[
level.repulsion
]
{\label{level.repulsion}
The two lowest $0^{-}$ \cfgs\ in \Pb\ are separated from the next
higher \cfgs\ by a large gap $\Delta$ allowing to discuss the simple
case of a two-level \cfg\ mixing in the $|5280\,0^{-}>$ and
$|5599\,0^{-}>$ states.  The residual interaction is decomposed into
the m.e. $v_{11}$ and $v_{22}$ describing the shift of the two levels,
and the m.e. $v_{12}=v_{21}$ describing the level repulsion.
}
\resizebox{\hsize}{09.51cm}{
{\includegraphics[angle=00]{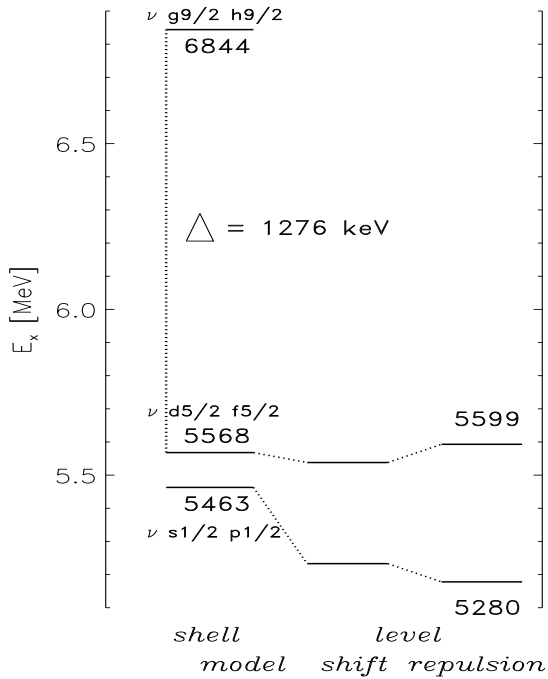}}
}
\end{figure}

\subsection{Completeness of the strength in the truncated \cfg\ space
}

Higher $0^{-}$ states are not known, but they should have energies
above $E_x\approx6.8$\,MeV, see Fig.~\ref{level.repulsion}.  In
contrast, nine $1^{-}$ states are known as predicted by the shell
model.

The cross sections $\left<{\frac{d\sigma}{d\Omega}}(E_x)\right>$
(Table~\ref{tab.compare}) for the two $0^{-}$ states and all $1^{-}$
states up to $E_x=6.5$\,MeV are consistent with the data of
Refs.~\cite{Valn2001,ValnThesis} within the errors.  The ratios agree
also with the population strengths of Ref.~\cite{Schr1997} but they
are more precise.

The reaction \PbS{\it(d,~p)} excites the two \cfgs\ \sOhlb\pOhlb\ and
\dThlb\pOhlb\ in all  $1^{-}$ states, but only the \cfg\ \sOhlb\pOhlb\
in the $0^{-}$ states. The two lowest $0^{-}$ states contain almost
the complete \sOhlb\pOhlb\ $0^{-}$ strength by comparison to DWBA
calculations \cite{Valn2001,ValnThesis}.  Because higher \cfgs\ admix
little due to the gap $\Delta$ between the second and third $0^{-}$
\cfgs, \dFhlb\fFhlb\ and \gNhlb\hNhlb, being larger than ten times the
mean m.e., the deviation matrix $d$ almost vanishes.  By comparing
the detected strength of the $0^{-}$ and $1^{-}$ \sOhlb\pOhlb\ \cfgs,
we deduce an upper limit for $|d|$.

The 5292 $1^{-}$ state contains less than 90\% of the \sOhlb\pOhlb\
strength, since the ratio of its cross section to the sum of the two
$0^{-}$ states is less than the ratio 3:1 expected from the spin
factor ($2I+1$) [Eq.~(\ref{eq.spinFactor})].
Other $1^{-}$ states contain the remaining \sOhlb\pOhlb\ strength, but
the 5292 $1^{-}$ state contains also some of the \dThlb\pOhlb\
strength (besides other \cfgs\ not detected by \PbS{\it(d,~p)} but by
IAR-pp').  The missing \sOhlb\pOhlb\ strength is contained in the
other eight $1^{-}$ states.

(a)~All $1^{-}$ states except for the 5292 $1^{-}$ state listed in
Table~\ref{tab.compare} have rather flat \angDs\ for $\Theta=20^\circ-
30^\circ$.
For the states considered, the dependence of the cross section on the
energy $E_x$ for states with the same \cfg\ mixture is negligible
\cite{Valn2001,ValnThesis}.  
(b)~For the 5924 $2^{-}$ and 5947 $1^{-}$ states, the \angD\ for
$\Theta=20^\circ- 30^\circ$ is flat (similarly as for states with
\dFhlb\pOhlb\ strength) in contrast to the steep rise for the
\sOhlb\pOhlb\ \cfg\ \cite{Valn2001,ValnThesis}.
The 5924 $2^{-}$ and 5947 $1^{-}$ states contain most of the
\dThlb\pOhlb\ strength \cite{Valn2001,ValnThesis} and the spin
assignments are firm \cite{Schr1997}.
(c)~In the 5947 state, the comparison of the shape of the \angD\ to
the 5924 $2^{-}$ state allows to deduce an upper limit for the
\sOhlb\pOhlb\ strength of about 8\% or a ratio $r_{2:0}= t_{1\,n3}^2/
t_{1\,n1}^2 >12$ [Eq.~(\ref{eq.ampl.1m})].
(d)~The deviation of the slope of the cross section for the 5292
$1^{-}$ state in comparison to the two $0^{-}$ states implies up to
10\% \dThlb\pOhlb\ admixture (Fig.~\ref{S.F.}).
(e)~For the other $1^{-}$ states besides the 5292 and 5947 states,
from the comparison of the shape of the \angD\ to the 5292 $1^{-}$ and
5924 $2^{-}$ states the ratio $r_{2:0}$ is derived, see
Table~\ref{tab.compare}.

Summing up thus derived upper limits of \sOhlb\pOhlb\ admixtures
$t_{1\,n1}^2$ to all other $1^{-}$ states, we derive a lower limit
80\% of the \sOhlb\pOhlb\ \cfg\ in the 5292 $1^{-}$ state.

Together with the upper limit of 90\% derived before, from
Eq.~\ref{eq.spinFactor} we conclude the sum of the \sOhlb\pOhlb\
strength in the 5280 $0^{-}$ and 5599 $0^{-}$ states to be complete
within better than 97\%.  This yields an upper limit for the deviation
matrix [Eq.~(\ref{eq.devia})],
\begin{eqnarray}
\label{eq.devWert}
d_{11} \approx d_{22} < 0.03,
\nonumber\\
|d_{12}| \approx |d_{21}| < 0.02.
\end{eqnarray}

\subsection{Excitation energies
}

From the known single particle and single hole states in the lead
region \cite{NDS1986}, the lowest one-particle one-hole \cfgs\ in \Pb\
with spin $0^{-}$ are predicted as
 $\nu$~\sOhlb\pOhlb,
 $\nu$~\dFhlb\fFhlb,
 $\nu$~\gNhlb\hNhlb,
 $\nu$~\dThlb\pThlb,
 $\pi$~\pThlb\dThlb\ (the lowest proton \ph\ \cfg)
at $E_x=
5463,
5568,
6844,
6866,
7383$\,keV, respectively, see Fig.~\ref{level.repulsion}.
%
The gap $\Delta$ described by Ref.~\cite{AB1973} between the two
lowest \cfgs\ \sOhlb\pOhlb\ and \dFhlb\fFhlb\ and the next \cfgs\ is
1276\,keV.  Since it is more than ten times higher than the mean
m.e. the mixing of the two lowest $0^{-}$ \cfgs\ in \Pb\ represents an
excellent example of a two-level scheme.

The energies of the shell model \cfgs\ are derived from the single
particle and single hole states in the four neighboring nuclei
\cite{NDS1986}, 
$ e^0=
\left(
\begin{array}{rr}
5463  & 0 \\
0  & 5568 \\
\end{array}
\right)
$ \,keV. The \expt al data yield the excitation energies of the two
states,
$ E=
\left(
\begin{array}{rr}
5280  & 0 \\
0  & 5599 \\
\end{array}
\right)
$ \,keV.

\begin{widetext}
\setlength\LTleft{0pt}
\setlength\LTright{0pt}
\begin{longtable*}
{@{\extracolsep{10pt}}
%
{l}@{}{c}
{c}@{}{c}
{c}@{}{c}
{c}@{}{c}
}
\caption{%
Levels near the 5280 $0^{-}$ and 5599 $0^{-}$ states in \Pb\
(marked~$\bullet$).  Within 1-2\,keV, the energy label corresponds to
the energies from
Refs.~\cite{Rad1996,1997RA207,Schr1997,Valn2001,ValnThesis,Yates96c}
or this work.  The values from Refs.~\cite{Valn2001,ValnThesis} refer
to the reaction \Pb{\it (p,~p'\,)} at $E_p=22$\,MeV.  Spin and parity
$I^\pi$ from Refs.~\cite{Schr1997,P.i11,Rad1996,1997RA207,Yates96c}
are shown.
}
\label{tab.0m}
\endfirsthead
\multicolumn{08}{c}
{Table I continued \dots}\\
\hline\hline
\endhead
\multicolumn{08}{c}{}\\
\endfoot
\multicolumn{08}{c}{}\\
\endlastfoot
\hline
\hline
&energy& $E_x$ & $E_x$ & $E_x$ & $E_x$ & $I^\pi$ & Ref.\\%
&label & keV& keV& keV& keV   &  & \\%
&      &{\it this work}& Ref.~\cite{Schr1997}&
Refs.~\cite{Rad1996,1997RA207}& Refs.~\cite{Valn2001,ValnThesis}& & \\%
\hline
\multicolumn{ 8}{|c|}{region near 5280 $0^{-}$ and  5292 $1^{-}$}\\
\hline
&{     5239} & $ 5239.5\pm0.8$&$5239.35\pm0.36$&
 $              $  &{$              $}&   4$^{-}$  &\cite{P.i11} \\%
&{     5241} & $             $&$5241\hfill    $&
 $              $  &{$5240.8\pm1.5  $}&   0$^{+}$  &\cite{Yates96c} \\%
&{     5245} & $ 5245.4\pm0.3$&$5245.28\pm0.06$&
 $5245.2 \pm0.1 $  &{$5244.6 \pm1.0$}&   3$^{-}$  &\cite{Schr1997} \\%
&{     5254} & $ 5254.2\pm0.8$&$5254.16\pm0.15$&
 $              $  &{$              $}&       & \\%
&{     5261} & $ 5261.2\pm0.8$&$              $&
 $              $  &{$              $}&       & \\%
&{     5266} & $ 5266.6\pm0.9$&$              $&
 $              $  &{$              $}&       & \\%
 &{     5276} & $ 5276.3\pm0.4$&$              $&
  $              $  &{$5277.1\pm1.5  $}&   4$^{-}$  &\cite{P.i11} \\%
$\bullet$&
 {5280} & $ 5280.5\pm0.1$& $5280.32\pm0.08$&
 $5280.5 \pm0.1 $  &{$5281.3\pm1.5  $}&   0$^{-}$  &\cite{Schr1997} \\%
&{     5287} & $5287.8\pm1.9 $& $              $&
 $              $  &{$5287.2\pm1.5  $}&       & \\%
&{     5292} & $ 5292.2\pm0.1$& $5292.00\pm0.20$&
 $5292.1 \pm0.1 $  &{$5292.6\pm1.5  $}&   1$^{-}$  &\cite{Schr1997} \\%
&{     5307} & $5307.6\pm1.5 $& $              $&
 $              $  &{$              $}&            & \\%
&{     5316} & $5313.0\pm1.0 $& $5317.00\pm0.20$&
 $              $  &{$              $}&   ($3^{+}$)&\cite{Schr1997} \\%
&{     5317} & $5316.9\pm1.5$& $5317.30\pm0.06$&
 $              $  &{$5317.7\pm0.6  $}&       & \\%
&{     5326} & $             $& $              $&
 $              $  &{$5326.9\pm0.6  $}&            & \\%
&{     5339} & $5340.0\pm0.9$& $5339.46\pm0.16$&
 $              $  &{$5340.1\pm1.5  $}&  8$^{+}$ &\cite{Schr1997} \\%
&{     5347} & $ 5347.4\pm0.2$& $5347.15\pm0.25$&
 $              $  &{$5348.4\pm0.6  $}&   3$^{-}$  &\cite{Schr1997} \\%
\hline
\multicolumn{ 8}{|c|}{region near 5599 $0^{-}$}\\
\hline
&{     5548} & $  5548.5\pm0.4$& $5548.08\pm0.20$&
 $5548.2 \pm0.1 $ &{$5547.5\pm1.5  $}&   2$^{-}$ &\cite{Schr1997} \\%
&{     5557} & $5557.2\pm1.0  $& $              $&
 $              $ &{$5554.0\pm2.0  $}&      & \\%
&{     5563} & $  5563.9\pm0.3$& $5563.58\pm0.14$&
 $5563.6 \pm0.1 $ &{$5564.7\pm0.6  $}&   $3^{-},4^{-}$ &\cite{Schr1997} \\%
&{     5566} & $              $& $5566.00\pm0.60$&
 $              $ &{$              $}&   4$^{-}$ &\cite{Schr1997} \\%
&{     5572} & $  5572.0\pm0.8$& $              $&
 $              $ &{$              $}&      & \\%
&{     5577} & $5579.0\pm0.9  $& $              $&
 $              $ &{$5576.6\pm1.5  $}&      & \\%
&{     5587} & $  5587.4\pm1.0$& $              $&
 $              $ &{$5587.7\pm0.5  $}&      & \\%
$\bullet$&
 {     5599} & $  5599.8\pm0.5$& $5599.40\pm0.08$&
 $5601.7 \pm0.1 $ &{$5599.6\pm0.4  $}&   0$^{-}$ &\cite{Schr1997} \\%
&{     5614} & $5614.4\pm1.7  $& $              $&
 $              $ &{$5615.4\pm0.4  $}&           & \\%
&{     5641} & $  5640.7\pm0.6$& $5641.10\pm0.50$&
 $5641.4 \pm0.5 $ &{$5639.9\pm1.5  $}&($1^{-},2^{+}$)&\cite{Rad1996,1997RA207} \\%
&{     5643} & $              $& $              $&
 $              $ &{$5643.1\pm1.5  $}&           & \\%
&{     5649} & $  5648.7\pm0.5$& $5649.70\pm0.28$&
 $              $ &{$5649.8\pm0.9  $}&  (5$^{-}$)& \\%
\hline
\hline
\end{longtable*}
\end{widetext}

\section{Results and Discussion
}
\subsection{Determination of matrix elements of the residual
interaction
}

The matrix elements of the residual interaction between the two lowest
$0^{-}$ \cfgs\ are derived in the truncated space of the first two
\cfgs\ by the method described in Ref.~\cite{AB1973},
\begin{eqnarray}
\label{eq.general}
v=tEt^\dagger - {\scriptstyle\frac{1}{2}}(tt^\dagger e^0 + e^0
tt^\dagger) + r, 
\end{eqnarray}
where $r$ is the residual matrix describing the influence of the
higher \cfgs\ $|C_q>$ in the space separated from the two lowest
\cfgs\ by the gap $\Delta$ (Fig.~\ref{level.repulsion}).

Explicitly we have
\begin{eqnarray}
\label{eq.v11v22v12}
v_{11} = t_{11}^2 E_{11} +  t_{12}^2 E_{22} -
(t_{11}^2+ t_{12}^2) e_{11}^0 + r_{11},
\nonumber\\
v_{22} = t_{21}^2 E_{11} + t_{22}^2 E_{22} -
(t_{21}^2+ t_{22}^2)  e_{22}^0 + r_{22},
\nonumber\\
v_{12} = t_{11} t_{21} E_{11} + t_{12} t_{22} E_{22} -
\nonumber\\
{\scriptstyle\frac{1}{2}}
(t_{11} t_{21} + t_{12} t_{22}) (e_{11}^0 +e_{22}^0) + r_{12}.
\nonumber\\
v_{21} = t_{21} t_{11} E_{11} + t_{22} t_{12} E_{22} -
\nonumber\\
{\scriptstyle\frac{1}{2}}
(t_{21} t_{11} + t_{22} t_{12}) (e_{11}^0 +e_{22}^0) + r_{21}.
\end{eqnarray}

Using Eqs. (\ref{eq.ampl}, \ref{eq.devWert}, \ref{eq.v11v22v12}) we
obtain the m.e.
\begin{eqnarray}
\label{eq.result}
v_{11}=-140\pm10\,(exp.)\,\pm40\,(syst.)\, {\rm keV},
\nonumber\\
\, v_{22}=- 5\pm10\,(exp.)\,\pm40\,(syst.)\, {\rm keV}, 
\nonumber\\
\, v_{12}= v_{21}= \pm(105 \pm10)\,(exp.)\,\pm40\,(syst.)\, {\rm
keV}.
\end{eqnarray}
The sign of the off-diagonal terms $v_{12},v_{21}$ cannot be
determined from our data.
The diagonal terms $v_{11}, v_{22}$ describe the level shift, the
off-diagonal terms $v_{12}, v_{21}$ the level repulsion, see
Fig.~\ref{level.repulsion}.

The m.e. (especially the off-diagonal m.e.) agree with the mean
m.e. of about 100\,keV obtained from the analysis of the lowest 20
\ph\ \cfgs\ in \Pb, see \cite{AB1973,Heu2006}.  The values $v$ are
compatible with theoretical calculations \cite{2000Brwn,Rej1999}, but
more precise.

The systematic error is well estimated for the diagonal
m.e. \cite{AB1973} by use of the deviation matrix $d$
[Eq.~(\ref{eq.devWert})].
The systematic error for the off-diagonal m.e. is estimated from the
residual matrix element
\begin{eqnarray}
\label{eq.off-rest}
r_{12}= \sum_q(t_{11}E_{11}t_{q1} + t_{1q}E_{qq}t_{11}).
\end{eqnarray}
From Eqs. (\ref{eq.unitary},~\ref{eq.devWert}) we derive contributions
from higher states and higher \cfgs\ to be small, $|t_{1q}|<0.14,
|t_{q1}|<0.14$.  Shell model calculations support the assumption of
statistically distributed signs for the amplitudes $t_{1q},t_{q1}$.
So, a systematic error of the off-diagonal m.e. of about 40\,keV may
be assumed.

\subsection{Data from IAR-pp'
}

A preliminary analysis of the IAR-pp' data \cite{P.i11} is consistent
with the spin assignments given in Table~\ref{tab.compare}.
Especially the 5292 $1^{-}$, 5924 $2^{-}$, 5947 $1^{-}$ states are
selectively excited by the \sOhlb, \dThlb, \dThlb\ IAR, respectively.

In early IAR-pp' \expt s \cite{1968WH02} \excF\ were measured for
several multiplets with a resolution of 26\,keV.  The energies given
by Ref.~\cite{1968WH02} derive from the calibration of IAR-pp' spectra
taken with the Enge split-pole magnetic spectrograph \cite{1967MO25}.
They are about 0.13\% too low \cite{P.i11}.

Measurements of the \excF\ for the unresolved 5280 $0^{-}$, 5292
$1^{-}$ doublet (``5.284 MeV'') show  a strong excitation by the
\sOhlb\ IAR.  A weak excitation by the \dFhlb\ IAR is explained by the
\dFhlb\fFhlb\ component in the 5280 $0^{-}$ state
[Eqs. (\ref{eq.ampl.0m},~\ref{eq.ampl})] and \dFhlb\fFhlb,
\dFhlb\pThlb\ components in the 5292 $1^{-}$ state
[Eq.\ref{eq.ampl.1m}].

Similarly the resolved 5924 $2^{-}$, 5947 $1^{-}$ doublet (``5.914 +
5.936 MeV'') is dominantly excited by the \dThlb\ IAR proving the
presence of about equal \dThlb\pOhlb\ components in both states in
agreement with the results from \PbS{\it(d,~p)}.  Whereas the 5924
state clearly resonates on the \sOhlb\ IAR (which is explained by weak
\sOhlb\fFhlb\ and \sOhlb\pThlb\ components), the decay curve of 5947
state near the \sOhlb\ IAR is smooth.

The \dFhlb\ and \sOhlb\ IAR are not well isolated, $E^{res}=16.496,
16.965$\,MeV and $E^{tot}=45\pm5, 45\pm8$, respectively
\cite{1968WH02,P.i11}.  Assuming isolated IAR and using the amplitudes
of Eq.~(\ref{eq.result}), a calculation of the cross sections for the
5280 $0^{-}$ and 5599 $0^{-}$ states on the \dFhlb\ and \sOhlb\ IAR
(using the IAR parameters of Ref.~\cite{P.i11}) roughly agrees with
the measured data.  An essay following Ref.~\cite{Heu1969a} to
describe the \angDs\ by interfering IAR did not yield conclusive
results essentially because of missing data at scattering angles
$\Theta<40^\circ$.

\section{Summary
}

Up to $E_x=6.1$\,MeV, the shell model predicts about 120 one-particle
one-hole states in \Pb\ but only two states with spin $0^{-}$.  From a
measurement of the reaction \PbS{\it(d,~p)} at a resolution of 3\,keV,
we identify the two known states with spin $0^{-}$ among about 150
states in a region where the mean level spacing is 6\,keV.
Spectroscopic information for the two $0^{-}$ states is used to
determine their structure.
 
Matrix elements of the residual interaction for the unique case of a
two-level mixing between the two lowest $0^{-}$ \cfgs\ in \Pb\ are
derived with higher precision than current shell model calculations.
Spectroscopic information for the nine lowest $1^{-}$ states is used
to quantify the systematic uncertainty.

Additional data from inelastic proton scattering via IAR in \Bi\
support the structure information obtained.

\acknowledgments{
This work has been supported by MLL, DFG C4-Gr894/2-3, and DFG
Br799/12-1.
}

\begin{widetext}
\setlength\LTleft{0pt}
\setlength\LTright{0pt}
\begin{longtable*}
{@{\extracolsep{00pt}}
%
                     {c}
@{\extracolsep{15pt}}{c}
@{\extracolsep{15pt}}{c}
@{\extracolsep{15pt}}{c}
@{\extracolsep{05pt}}{r}
@{\extracolsep{00pt}$~ \pm$\extracolsep{00pt}}{c}
@{\extracolsep{05pt}}{r}
@{\extracolsep{15pt}}{c}
@{\extracolsep{05pt}}{r}
@{\extracolsep{15pt}}{c}
@{\extracolsep{00pt}}{c}
@{\extracolsep{05pt}}{r}
@{\extracolsep{02pt}$~ \pm$\extracolsep{00pt}}{c}
@{\extracolsep{02pt}}{r}
}
\caption{%
Up to $E_x=6.5$\,MeV, for the two states with spin $0^{-}$
(marked~$\bullet$) and nine states with spin $1^{-}$, the mean
cross section $\left<{\frac{d\sigma}{d\Omega}}(E_x)\right>$ [see
Eq.~(\ref{eq.mean})] adjusted to reproduce the cross section at
$\Theta=25^\circ$ for the 5292 $1^{-}$ state is shown.  Within
1-2\,keV, the energy label reflects the energies $E_x$ from
Refs.~\cite{Rad1996,1997RA207,Schr1997,Valn2001,ValnThesis} or this
work.
Spectroscopic factors $S_{(d,p\gamma)}$ \cite{Schr1997} and
S.F. \cite{Valn2001,ValnThesis} are given for comparison. The reaction
\PbS{\it(d,~p)} was measured with the same deuteron energy
$E_d=22.000$\,MeV as Refs.~\cite{Valn2001,ValnThesis}.
In the states with spin $1^{-}$, the $L=0$ and $L=2$ transfer excites
the \sOhlb\pOhlb\ and \dThlb\pOhlb\ \cfgs, respectively, but in the
two $0^{-}$ states only the \sOhlb\pOhlb\ component is excited by the
$L=0$ transfer [Eqs. (\ref{eq.ampl.0m}), (\ref{eq.ampl.1m})].  From
the measured \angDs, we derive the ratio $r_{2:0}$ of the strength
$t^2$ for the \cfgs\ \dThlb\pOhlb\ ($L=2$) and \sOhlb\pOhlb\
($L=0$). Namely, the \angD\ for $L=2$ is flat in contrast to the steep
slope for $L=0$.  For the same S.F. the relative cross section at
$\Theta=25^\circ$ $\left<{\frac{d\sigma}{d\Omega}}(E_x)\right>$ rates
as about $1:0.5$ for $L=2$ to $L=0$ \cite{Valn2001,ValnThesis}.
}
\label{tab.compare}
\endfirsthead
\multicolumn{14}{c}
{Table II continued \dots}\\
\hline\hline
\endhead
\multicolumn{14}{c}{}\\
\endfoot
\multicolumn{14}{c}{}\\
\endlastfoot
%
\hline
\hline
 \multicolumn{1}{c}{$n$}
&\multicolumn{1}{c}{Energy}
&\multicolumn{1}{c}{$I^\pi$}
&\multicolumn{1}{c}{$L$}
&\multicolumn{3}{c}{$S_{(d,p\gamma)}$}
&\multicolumn{1}{c}{$L$}
&\multicolumn{1}{c}{S.F.}
&\multicolumn{2}{c}{$r_{2:0}$}
&\multicolumn{3}{c}{$\left<{\frac{d\sigma}{d\Omega}}(E_x)\right>$}
\\%
&label
&
&
&\multicolumn{3}{c}{$\times\,1000$}
&
&\multicolumn{1}{c}{$\times\,1000$}
&\multicolumn{2}{c}{}
&\multicolumn{3}{c}{$\mu b/sr$}
\\%
&
&\multicolumn{5}{c}{\_\_\_\_\_\_\_\_\_\_\_\_\_\_\_\_\_\_\_\_\_\_\_\_\_\_\_\_}
&\multicolumn{2}{c}{\_\_\_\_\_\_\_\_\_\_\_\_\_\_}
&\multicolumn{5}{c}{\_\_\_\_\_\_\_\_\_\_\_\_\_\_\_\_\_\_\_}
\\%
&
&\multicolumn{5}{c}{Ref.~\cite{Schr1997}}
&\multicolumn{2}{r}{Refs.~\cite{Valn2001,ValnThesis}}
&\multicolumn{5}{c}{this work}\\
\hline
 1&4841 &$1^{-}$& 0&   11&&  4 &  &     &$ >$&0.5    & 22&&  5\\%
$\bullet$ &5280 &$0^{-}$& 0&  377&& 32 & 0&  650&    & 0     &250&& 10\\%
 2&5292 &$1^{-}$& 0& 1071&& 325& 0& 1550&$ <$&0.1    &785&& 30\\%
 3&5512 &$1^{-}$& 0&   74&& 22 & ~&  ~~~&$ >$&0.8    &160&& 15\\%
\multicolumn{7}{c}{ }& 2&  165\\%
$\bullet$&5599 &$0^{-}$& 0&   60&&  6 & 0&  103& & 0&  40&&  5\\%
 4&5641 &$1^{-}$\footnotemark[1]\footnotetext[1]%
{$I^\pi=(1^{-},2^{+})$
from Refs.~\cite{Rad1996,1997RA207}. The prelimary analysis of our data excludes spin $2^{+}$.}
&&\multicolumn{1}{r}{ 4 \footnotemark[2]
\footnotetext[2]%
{Derived from the relative population strength ($S_{expt}$).}
}
&& && &$>$&0.7 &  22&&  3\\%
 5&5947 &$1^{-}$& 2&{ 1266}&&{488 }& 2&1390&
  $>$&12\footnotemark[3]\footnotetext[3]%
{By comparison to the 5924 $2^{-}$ state with $L=2$ only.}
  & 1300&&    80\footnotemark[4]\footnotetext[4]%
{The error includes the variation of the \angD\ with~$\Theta$.}\\%
 6&6263 &$1^{-}$& 2&{   55}&&{ 23 }& 2&    7&$>$&0.6    &  25&& 10\\%
\multicolumn{7}{c}{ }& 0&   59\\
 7&6314 &$1^{-}$& 2&{   88}&& 38& 0&  113&$>$&0.7 &  38&& 12\\%
 8&6360 &$1^{-}$& 2&{   29}&& 13& 2&   13&$>$&0.7 &   9&&  3\\%
 9&6486 &$1^{-}$\footnotemark[5]\footnotetext[5]%
{$I^\pi=1^{-}$ from Refs.~\cite{Rad1996,1997RA207}.}
&&\multicolumn{1}{r}{ 30\footnotemark[2]
}
&\multicolumn{1}{c}{}
&& 2&   38&$>$&0.8 & 30&&  5\\%
\multicolumn{7}{c}{ }& 0&   12\\%
\hline
\hline
\end{longtable*}
\end{widetext}
%

%
\end{document}